\documentstyle[12pt,epsfig]{article}
\begin{document}
\newcommand{\beq}{\begin{equation}}
\newcommand{\eeq}{\end{equation}}
\topmargin-1cm
\oddsidemargin0pt
\evensidemargin0pt
\textheight41\baselineskip
\textwidth16cm
%
\noindent
\hspace*{10.5cm}Dec. 1995\\
\hspace*{10.5cm}OU-HET 228\\
\hspace*{10.5cm}TOYAMA - 86\\
\hspace*{10.5cm}UT-DP-95-01\\
\hspace*{10.5cm}hep-ph/9512387

\begin{center}
  {\LARGE $W_R$ effects on $CP$ angles determination \\
at a $B$ factory}
\end{center}
\vspace*{0.1cm}
\begin{center}
{\large T.~Kurimoto}\footnote{e-mail: krmt@sci.toyama-u.ac.jp}\\
Department of Physics, Faculty of Science,\\
Toyama University,\\
Toyama 930, Japan\\
\vspace*{.2cm}
{\large A.~Tomita}\footnote{e-mail: tomita@phys.wani.osaka-u.ac.jp
(available until March 1996)}\\
Department of Physics, Faculty of Science,\\
Osaka University,\\
Toyonaka, Osaka 560, Japan\\
\vspace*{.2cm}
{\large S.~Wakaizumi}\footnote{e-mail: wakaizumi@medsci.tokushima-u.ac.jp}\\
Department of Physics, School of Medical Sciences,\\
University of Tokushima,\\
Tokushima 770, Japan\\

\vspace*{0.5cm}
\Large{\bf Abstract}
\end{center}
The right-handed charged current gauge boson
$W_R$ can affect significantly on the determination of the $CP$ violation
angles to be measured at $B$ factories if the right-handed current
quark mixing matrix $V^R$ is taken to a specific form to
satisfy the bounds by neutral $K$ meson systems. The
$W_R$ contribution can be sizable in $B^0$-$\overline{B^0}$ mixing and
tree level $b$ quark decay. The deviation of $CP$ angles
in unitarity triangle from the standard model values can be as large as
$-37^\circ$ or $ +22^\circ$ for $\phi_3$ ($\gamma$), and
$66^\circ \sim 115^\circ$  for
$\phi_1$ ($\beta$) and $\phi_2$ ($\alpha$).
\section{Introduction}
One of the main purposes of the $B$ factories which are now under
construction at KEK\cite{KEKB} and SLAC\cite{SLACB}
is to find a sign of new physics beyond the standard model.
If a new physics significantly affects $B^0$-$\overline{B^0}$ mixing or
$CP$ violation in $B$ decays\cite{unit}, the angles and the sides of
the so-called unitarity triangle deviate from the standard model values,
and the consistency of the triangle will be lost\cite{newb}.
The extensive measurements at $B$ factories are going to
fix all the sides and angles of the unitarity triangle. We can
find a signal of new physics or at least constraints
by over-checking the consistency of the triangle.

New physics effects are often considered to appear at loop level
because the masses of most of the new particles in the models beyond
the standard model
are larger than the electro-weak scale of $O(M_W)$ to satisfy the bounds
by the present experimental data.
A heavy new particle would not be able to give a significant
contribution to the tree level $b$ quark decay.
However, this is not always the case.
The decay of $b$ quark through the standard $W$ boson exchange is suppressed
by the smallness of the involving Kobayashi-Maskawa (KM) matrix\cite{KM}
elements, $V_{cb}$ and $V_{ub}$. A new particle can contribute
significantly to tree level $b$ decay if it has non-suppressed coupling
with quarks. We consider the $W_R$ boson
in the $SU(2)_L \times SU(2)_R \times
U(1)$ models\cite{lrmd} as an example of such a new particle in this paper.

In addition to the ordinary KM matrix there exists a flavor mixing
matrix also in the coupling between
right-handed quark currents and $W_R$ boson
in $SU(2)_L \times SU(2)_R \times
U(1)$ models, which shall be called as  $V^R$  while the usual
left-handed current KM matrix  as  $V^L$  hereafter. It has been shown
that the mass of  $W_R$ should be greater than 1.4 TeV to be consistent
with the experimental data of $K^0$-$\overline{K^0}$ mixing
if a model has manifest or pseudo manifest left-right
symmetry, i.e. $V^L=V^R$ or $V^L=(V^R)^*$, respectively\cite{mrcon}.
The $W_R$ boson cannot contribute significantly to tree level $b$ decay
with such a heavy mass and $V^R$. But right-handed charged current interaction
has not been observed yet, so the the form of $V^R$ is not restricted
to manifest or pseudo manifest type.
Olness and Ebel have shown that
the mass limit of $W_R$ can be lowered to 300 GeV by assuming
specific forms of $V^R$\cite{OE}. Langacker and Sankar have also made
a detailed analysis on $W_R$ mass limit, and come to a similar
conclusion that the lower limit of $W_R$ mass can be reduced by taking
the following forms of $V^R$\cite{LS};
\beq
   V^R_I= \left(
       \begin{array}{ccc}
           e^{i\omega} & 0 & 0\\
           0 & ce^{i\xi} & se^{i\sigma}\\
           0 & se^{i\varphi} & ce^{i\chi}
       \end{array} \right) , \qquad
   V^R_{II}= \left(
       \begin{array}{ccc}
           0 & e^{i\omega} & 0 \\
           ce^{i\xi} & 0 & se^{i\sigma} \\
           se^{i\varphi} & 0 & ce^{i\chi}
       \end{array} \right) ,
  \label{eqn:rkm}
\eeq
where $s=\sin\theta$ and $c=\cos\theta$
($0\le \theta \le 90^\circ$).\footnote{The elements $0$ in
these matrices may be $O(10^{-2})$. We take them $0$ for
the simplicity of discussion.}
Unitarity requires $\xi-\sigma=\varphi-\chi +\pi $.
We call the former type of $V^R$ as type I and the latter
as type II in the following discussion. London and Wyler
have pointed out that both types of $V^R$ lead to sizable
contributions to $CP$ violation in $K$ and $B$ systems\cite{LW}.
For example, the $CP$ asymmetry in $B \rightarrow J/\Psi K_s$ decay can
be significantly altered by the presence of $W_R$ mediated
$b \rightarrow c\bar cs$ tree decay with the $V^R$ of type I.

The aim of this paper is to make a detailed study
on the effects of $W_R$ on the determination of the three $CP$ angles
to be measured at the B factories. In particular, we show that
the measurement of the angle $\phi_3$ (or $\gamma$)\footnote{
There are two notations for the angles of unitarity triangle.
One given in ref.\cite{KEKB} and the another in ref.\cite{PDG}.
We take the notation of ref.\cite{KEKB} as $\beta$ is used to express
another quantity here.}
can receive a sizable effect by $W_R$ in the case of type II right-handed
Kobayashi-Maskawa matrix, $V^R$, even if $W_R$ is as heavy as about 1 TeV.
The $CP$ angle  $\phi_3$ (or $\gamma$) was considered to
be measured on the $B_s$ decay, $B_s \rightarrow \rho K_S$ in many of
the previous papers. But the
experiments at B factories at the first stage  will be
made on $\Upsilon (4S)$ which can not decay into $B_s$. The measurement
of the angle  $\phi_3$ (or $\gamma$) is to be made on
$B \rightarrow DK$ decays\cite{unit,gamma}. Our analysis
on the angle  $\phi_3$ (or $\gamma$) is based on this method.

The rest of this paper is organized as follows:
In section 2 we give constraints of the parameters from
$K$ and $B$ systems. In section 3 we estimate the $W_R$ contribution to the
$CP$ violation angles in decays of $b$ quark.
The final section is devoted to summary and discussion.
\vspace*{1cm}
  \begin{center}
    \epsfig{figure=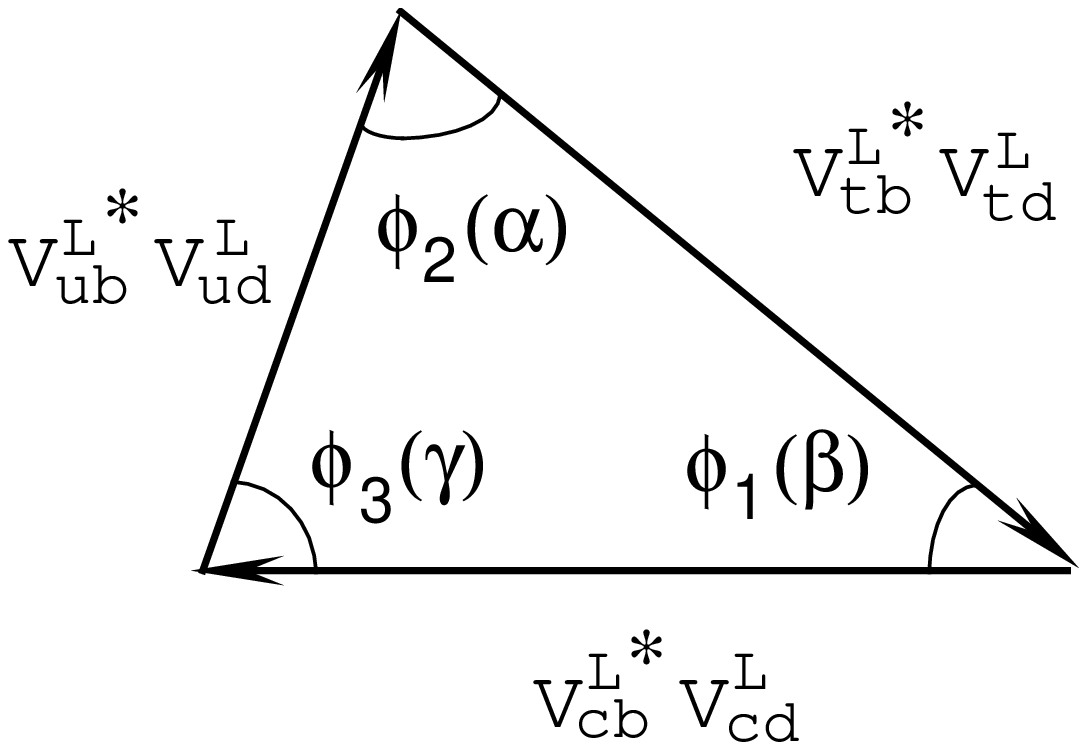, height=4cm}
\vspace*{4mm}

Figure 1: Unitarity triangle
  \end{center}
\vspace*{1cm}
%
\section{Constraints from $K$ and $B$ systems}
\subsection{$K^0$-$\overline{K^0}$ mixing}
The box diagram with one $W_L$ (standard $W$ boson) and one $W_R$
can give the major contribution
to $K^0$-$\overline{K^0}$ mixing in $SU(2)_L\times SU(2)_R \times U(1)$
models\cite{mrcon,ecg}:
\beq
{\cal H}^{eff}_{LR}
=\sum_{i,j=u}^t
\frac{2G_F^2M_W^2}{\pi^2}\beta_g
V^{L*}_{id}V^{R}_{is}V^{R*}_{jd}V^{L}_{js} J(x_i,x_j,\beta)\
\overline{d_R}s_L\overline{d_L}s_R + \mbox{(h.c.)},
\label{eqn:lrham}
\eeq
where $\beta$ is the square of the ratio of $W_L$ mass ($M_L$) to
$W_R$ mass ($M_R$), $M_L^2/M_R^2$, $\beta_g = (g_R/g_L)^2 \beta$ and
$x_i= m_i^2/M_L^2$. The loop function is defined as
\beq
J(x,y,\beta) \equiv \sqrt{xy}
     [(\eta^{(1)}+\eta^{(2)}\frac{xy\beta}{4})J_1(x,y,\beta)
        -\frac{1}{4}(\eta^{(3)}+\eta^{(4)}\beta)J_2(x,y,\beta)],
\label{eqn:lrh}
\eeq
with
\begin{eqnarray*}
  J_1(x,y,\beta) &=& \frac{x\ln x}{(1-x)(1-x\beta)(x-y)} +(x\leftrightarrow y)
     -\frac{\beta \ln \beta}{(1-\beta)(1-x\beta)(1-y\beta)} \ , \\
 J_2(x,y,\beta)&=&
\frac{x^2 \ln x}{(1-x)(1-x\beta)(x-y)} +(x\leftrightarrow y)
     -\frac{\ln \beta}{(1-\beta)(1-x\beta)(1-y\beta)} \ ,
\end{eqnarray*}
where $\eta^{(1)-(4)}$ are QCD corrections.
The box diagram with two $W_R$ cannot contribute to
$K^0$-$\overline{K^0}$ mixing as long as we take $V^R$ to be
in the forms of eq.(\ref{eqn:rkm}). We have neglected $W_L$-$W_R$ mixing as
it is highly suppressed by the experimental data\cite{LS}.
The real part of the matrix element
$\langle K^0 |{\cal H}^{eff}_{LR}|\overline{K^0}\rangle $ contributes to
$\Delta M_K$, while the imaginary part to the $CP$ violation parameter
$\epsilon$  in $K$ decay. The constraint by $\Delta M_K$ \cite{OE,LS}
is satisfied for $M_R> 0.52$ TeV which we take as
the limit from direct search\cite{PDG} under the assumption that right-handed
neutrino $\nu_R$ does not affect the b semi-leptonic decay.
The constraint from $\epsilon$ is much severe. $W_R$ has to be as heavy as
about 5 TeV or more unless the parameters in $V^R$ are tuned\cite{LW} .
By using the Wolfenstein parameterization\cite{wolf} of KM matrix $V^L$,
\beq
   V^L = \left(
       \begin{array}{ccc}
           1-\frac{\lambda^2}{2} & \lambda  & A\lambda^3 (\rho -i\eta)\\
           -\lambda & 1-\frac{\lambda^2}{2} & A\lambda^2 \\
           A\lambda^3 (1-\rho -i\eta)& - A\lambda^2 & 1
       \end{array} \right) ,
\label{eqn:wolf}
\eeq
we find that the contribution to $\epsilon$ has the following
combinations of quark mixing matrix elements:
\begin{description}
\item[type I] ($uc$ contribution): $\lambda^2 c \sin (\omega-\xi)$
\item[type I] ($ut$ contribution): $A \lambda^4 s[(1-\rho)\sin
                (\varphi-\omega)+\eta
                \cos(\varphi-\omega)]$
\item[type II] ($uc$ contribution):
      $(1-\frac{\lambda^2}{2})^2 c \sin (\omega-\xi)$
\item[type II] ($ut$ contribution): $-A\lambda^2 (1-\frac{\lambda^2}{2})
s \sin (\omega-\varphi)$
\end{description}
To suppress large contribution to $\epsilon$ from $W_L$-$W_R$ box diagram
we take the following solutions for $V^R$;
\begin{description}
\item[type I] \
  \beq
  \sin (\omega-\xi)=0  \mbox{ and } \tan (\omega-\varphi)= \frac{\eta}{1-\rho},
  \label{eqn:soli}
\eeq
\item[type II] \
\beq
 c =0 \mbox{ and } \sin (\omega-\varphi)=0.
\label{eqn:sol}
\eeq
\end{description}
There are other solutions to suppress $\epsilon$. We take these  since
they give most significant effects on $CP$ violation in $B$ decay by $W_R$.
\subsection{$B$ semi-leptonic decay}
$B$ semi-leptonic decays gives the constraint on the
ordinary KM matrix elements, $|V_{cb}|$
and  $|V_{ub}|$. It is independent of the $W_R$ effects as far as
the right handed neutrinos are heavier than $b$ quark,
which we take as a reasonable assumption. Then $B$ semi-leptonic decays
constrain only the elements of $V^L$.
\subsection{$B^0$-$\overline{B^0}$ mixing}
Let us write the contribution to $B^0$-$\overline{B^0}$ mixing matrix
elements as
\beq
M_{12}^B = M_{12}^{SM}[1 + d_{LR} + d_{RR}],
\label{eqn:mbb}
\eeq
where $d_{LR}$ and  $d_{RR}$ are the contributions by the box diagrams
with one $W_R$ and those with two $W_R$, respectively.
The contribution by $W_R$ depends on which of $V^R$ in type I and II we
take. In the following discussions
we take the gauge coupling
of $SU(2)_L$  ($g_L$) and that of $SU(2)_R$  ($g_R$)
equal for the simplicity of the following arguments.
$W_L$-$W_R$ mixing is neglected.
\subsubsection{Type I}
The contribution $d_{RR}$ vanishes because
$V^R_{Iib} V^{R*}_{Iid} =0$ for any $i=u,c,t$. By replacing $s$ with $b$
in eq.(\ref{eqn:lrham}) we can calculate $d_{LR}$, which is written as
\beq
 d_{LR} = \frac{V^L_{ub}V^{L*}_{cd}s e^{i(\sigma-\omega)}}{
                  (V^L_{tb}V^{L*}_{td})^2}r_{uc} +
             \frac{V^L_{ub}V^{L*}_{td}c e^{i(\chi-\omega)}}{
                  (V^L_{tb}V^{L*}_{td})^2}r_{ut}\  ,
\label{eqn:clr}
\eeq
where $r_{uc}$ ($r_{ut}$) are the ratios of the $W_R$ contribution
to the standard model one of
$u, c$ ($u, t$) quarks up to quark mixing matrix elements. We find
$|r_{uc}|$ is $O(10^{-6})$ or less and $|r_{ut}|$ is $O(10^{-5})$ or less
for $M(W_R) > 0.5$ TeV, while the absolute
magnitudes of the coefficients of $r_{uc}$ and $r_{ut}$ in
eq.(\ref{eqn:clr}) are  $O(10)$ and $O(1)$, respectively.
(Note that $|V^L_{td}|$ should be $O(\lambda^3)$ to realize
 the experimental value of $\epsilon$ in $K$ system.)
Therefore, we can neglect the $W_R$ contribution to
$B^0$-$\overline{B^0}$ mixing in the case of type I.
\subsubsection{Type II}
We take the solution given in eq.(\ref{eqn:sol}) to suppress the
large contribution to $\epsilon$. Then the contribution $d_{RR}$
vanishes. The contribution $d_{LR}$ is given
in the Wolfenstein parameterization of $V^L$ as,
\beq
d_{LR} =
      \frac{V^L_{tb}V^{L*}_{cd} e^{i(\sigma-\varphi)}}{
                  (V^L_{tb}V^{L*}_{td})^2}r_{ct}
     =
       -\frac{e^{i(\sigma-\varphi-\phi_{SM})} }{A^2\lambda^5
              [(1-\rho)^2 + \eta^2]}  r_{ct} ,
\label{eqn:dlr}
\eeq
where $\phi_{SM}=\arg (M_{12}^{SM})$. The ratio $r_{ct}$ is $O(10^{-3})$
as shown in Fig.2.
  \begin{center}
    \epsfig{figure=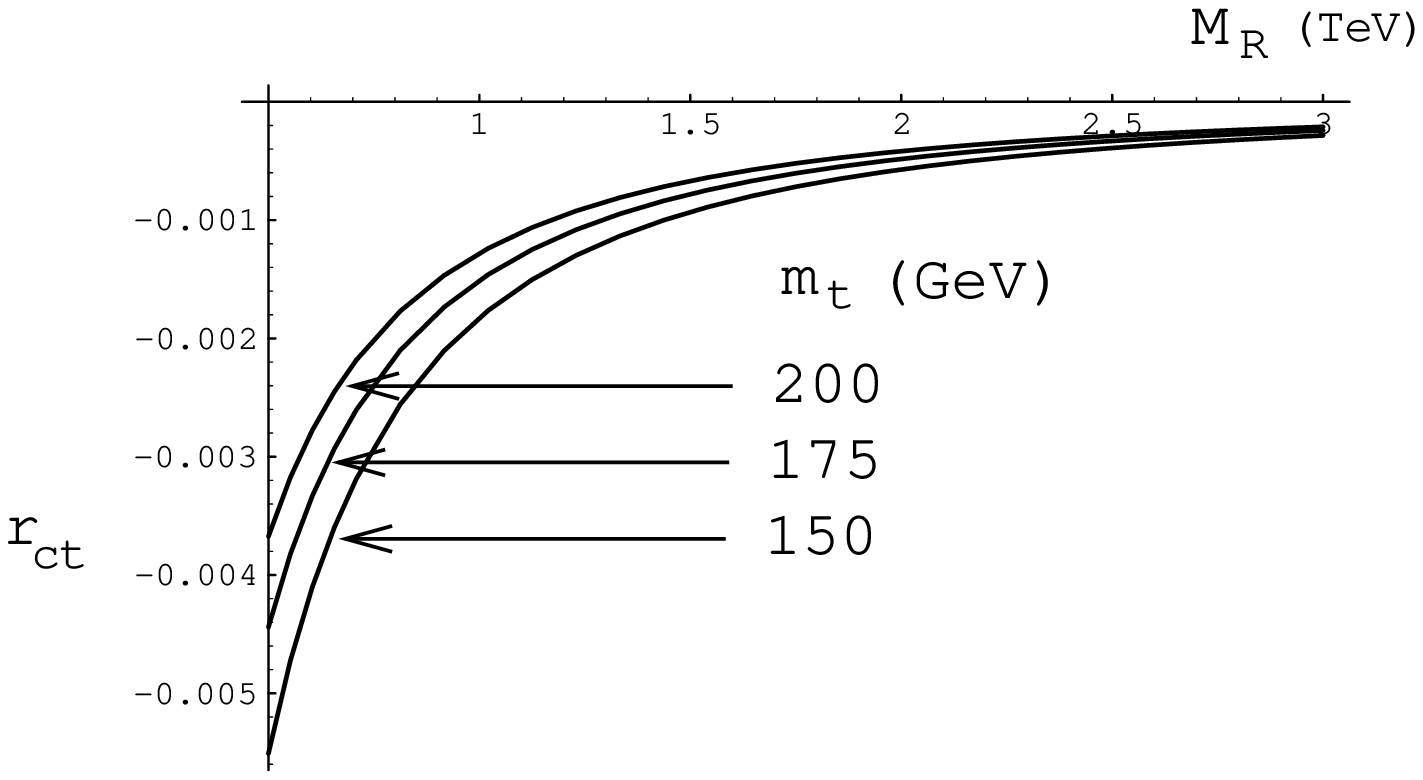,height=5cm}
\vspace*{4mm}

  Figure 2:
  \begin{minipage}[t]{10cm}
The ratio of $W_R$ contribution to $W_L$ contribution
up to quark mixing matrix elements.
  \end{minipage}
  \end{center}

\noindent
We have used
$\eta^{(1)}=1.1$, $\eta^{(2)}=0.26$,
$\eta^{(3)}=1.1$  $\eta^{(4)}=1.0$ for $W_R$ contribution\cite{ecg},
$\eta_{tt}=0.8$ for $W_L$ contribution\cite{datta}
as the values of QCD corrections, $m_c=1.5$ GeV and $m_b=4.6$ GeV
in calculating $r_{ct}$.The factor
$1/(A^2\lambda^5 [(1-\rho)^2 + \eta^2])$ is $O(10^{3})$.
The experimental value of $B^0$-$\overline{B^0}$ mixing can be realized
depending on the parameters in $V^L$ and $V^R$. We fix $\lambda=0.22$,
$A=0.8$, and investigate the
following two cases;
\beq
 r_B \equiv \sqrt{(1-\rho)^2 + \eta^2} =
\left\{
    \begin{array}[c]{lr}
     1.3 & \mbox{: case (a)} \\
     1.0 & \mbox{: case (b)}
    \end{array}. \right.
\eeq
The above two cases are shown in Fig.3  with the allowed regions
by $\epsilon$ in $K$ decay and $B$ semi-leptonic decay, where we take
$m_t= 150 \sim 200$ GeV and $B_K= 0.6 \sim 1.0$.
We find that the allowed region of $\phi_1$ ($\beta$) is
$7^\circ \sim 15^\circ$ for case (a), $13^\circ \sim 25^\circ$
for case (b).
The phase of $M_{12}^{SM}$, $\phi_{SM}$,
 is given by $\arg [(V^L_{tb}V^{L*}_{td})^2]
= 2 \phi_1$ in the phase convention of $V^L$ in eq.(\ref{eqn:wolf}).
  \begin{center}
    \epsfig{figure=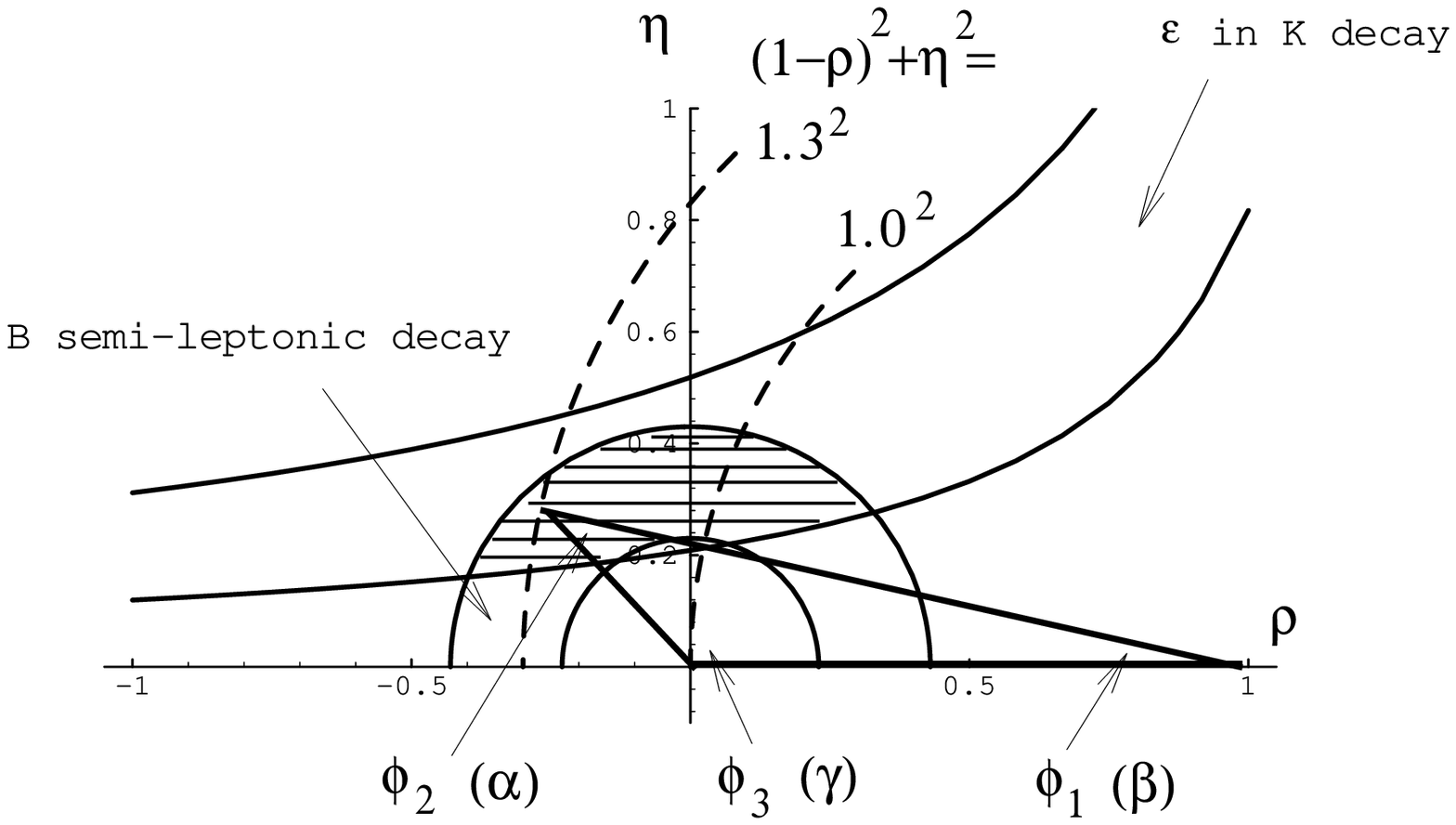,height=6.5cm}
\vspace*{4mm}

  Figure 3:
  \begin{minipage}[t]{10cm}
Allowed region in $\rho$-$\eta$ plane consistent with
     $\epsilon$ in $K$ decay and $B$ semi-leptonic decay. The dashed lines
     shows the curves $(1-\rho)^2 + \eta^2 = 1.3^2$ and $1.0^2$.
     The three angles of unitarity triangle are also shown.
  \end{minipage}
   \end{center}

\noindent
The ratio of the experimental data, $M_{12}^{B(exp)}$, for
 $\Delta M_B = 0.462 \pm 0.026 {\rm ps}^{-1}$ \cite{blow}
to the standard model contribution is given for $m_t=150,175,200$ GeV,
$r_B=1.3,1.0$, $f_B=140 \sim 200$ MeV and $B_B=0.7 \sim 1.1$ in Table 1:
  \begin{center}
  \begin{tabular}{c||ccc}
$r_B$ & $m_t$ = 150 (GeV) & 175 & 200 \\ \hline
1.3 & 0.42 $\sim$ 1.5 & 0.33 $\sim$ 1.2 & 0.27  $\sim$ 0.96 \\
1.0 & 0.70 $\sim$ 2.5 & 0.55 $\sim$ 2.0 & 0.45 $\sim$ 1.6
\end{tabular}
\vspace*{4mm}

Table 1 : $|M_{12}^{B(exp)}/M_{12}^{SM}|$
    \end{center}

\noindent
The equations (\ref{eqn:mbb}) and (\ref{eqn:dlr}) give
\beq
 \left| \frac{M_{12}^{B}}{M_{12}^{SM}}\right| =
     \left| 1 - e^{i(\sigma-\varphi-\phi_{SM})}
     \frac{r_{ct}}{A^2\lambda^5 [(1-\rho)^2 + \eta^2]}\right| .
\label
{eqn:res}
\eeq
We require the right-hand side of eq.(\ref{eqn:res}) should be
within the values give in Table 1. Taking as an example $m_t=175$ GeV,
$r_B=1.3$ and $M_R=1.2$ TeV, we obtain the allowed region of phases
in $V^R$ as seen in Fig.4.
It gives the range of the angle $\sigma-\varphi-\phi_{SM}
= 154^\circ \sim 206^\circ$, which gives the deviation of the $\arg M_{12}$
from the standard model value, $\delta\phi_{SM}=131^\circ \sim 229^\circ$.

There should be overlap between the shaded region and the circle whose
center is $(1,0)$ in Fig.4 for the consistency with the experimental value of
$\Delta M_B$. The radius of the circle depends on $W_R$ mass, so that
we can derive the bound of $M_R$, which is shown in Table 2.
  \begin{center}
  \begin{tabular}{c||ccc}
$r_B$ & $m_t$ = 150 (GeV) & 175 & 200 \\ \hline
1.3 & $>1.1$ & $>1.1$ &$ 11>M_R>1.0$ \\
1.0 & $>1.3$ & $>1.2$ &  $>1.2$
\end{tabular}
\vspace*{4mm}

 Table 2 :  $W_R$ mass bound (TeV)
    \end{center}

\noindent
Note that we need a new physics contribution to cancel too large
$W_L$ contribution in the case of $m_t=200$ GeV and $r_B=1.3$.
  \begin{center}
    \epsfig{figure=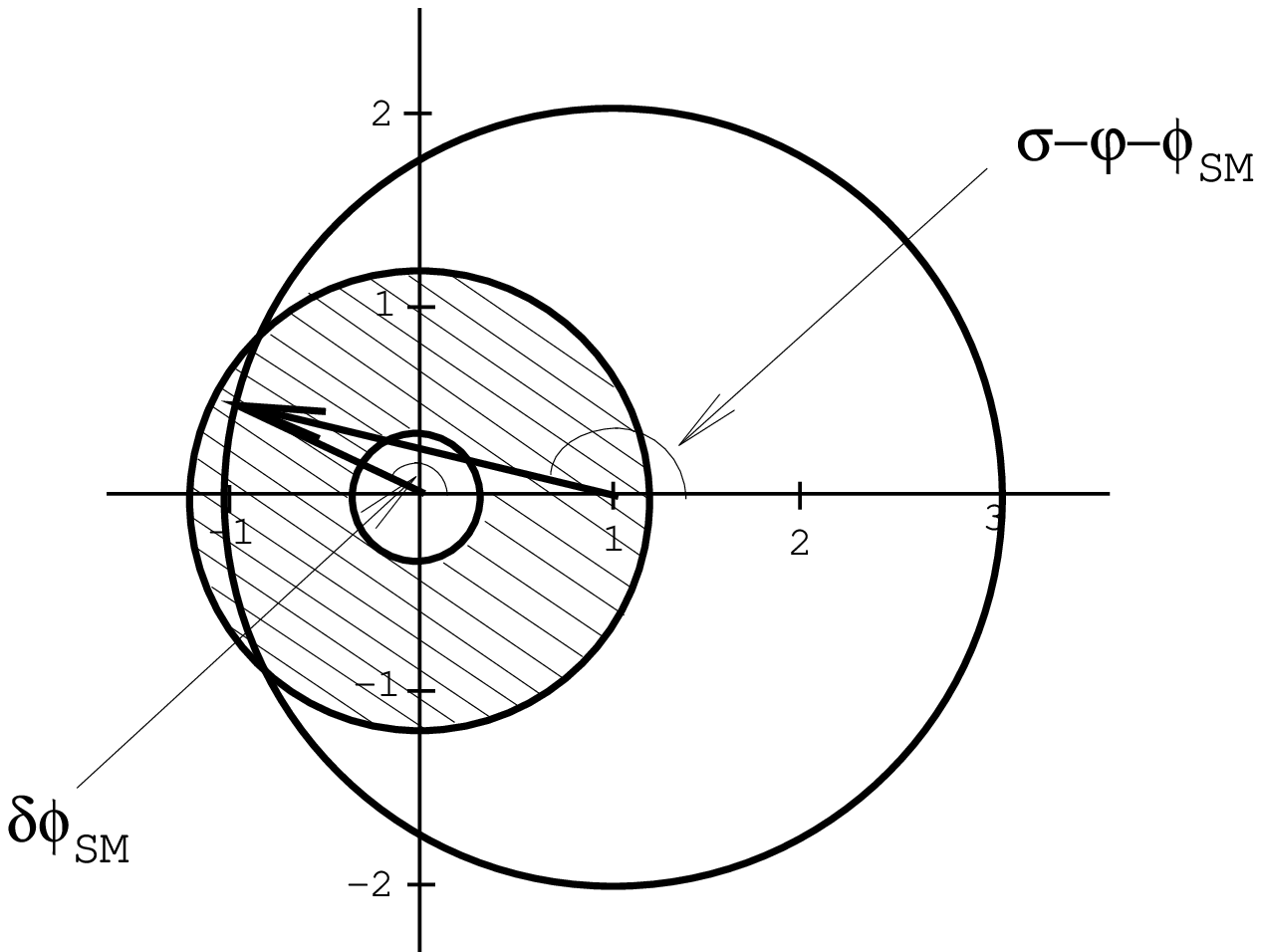,height=7cm}
\vspace*{4mm}

Figure 4:
\begin{minipage}[t]{10cm}
Allowed region of $d_{LR}$. The shadowed region corresponds to
the left-hand side of eq.(11) with the value given in Table 1.
The vector with arrow expresses the second term in the right-hand side of
eq.(11).
    \end{minipage}
   \end{center}
%
\section{Effects on $CP$ angles}
\subsection{$\phi_1$}
The $CP$ angle $\phi_1$ ($\beta$) is measured in the decay,
$B^0 (\overline{B^0}) \rightarrow J/\Psi K_S$.
It is $b(\bar b) \rightarrow c\bar c s(\bar s)$ decay
in the quark picture. The involving quark mixing matrix elements are
$V_{cb}$ and $V_{cs}$. The experiments fix the angle $\phi_1^{exp}$
which coincides with the angle $\phi_1$ of the unitarity triangle
in the case of standard model.
It is related to $M_{12}^B$ and the decay amplitudes
as follows:
\beq
\sin 2\phi_1^{exp}= -{\rm Im}\left [ \frac{|M_{12}^B|}{M_{12}^B}
     \frac{A(\overline{B^0}\rightarrow J/\Psi K_S )}
          {A(B^0 \rightarrow J/\Psi K_S)} \right].
\eeq
Penguin diagram also contributes to this decay in addition to
the tree graphs. The phase of $W_L$ ($W_R$) penguin amplitude has the same
phase with $W_L$ ($W_R$) tree amplitudes.
In the case of $V^R$ of type I  the ratio of $W_L$
contribution to $W_R$ contribution
in decay amplitudes is given as
\begin{eqnarray}
   W_L \ {\rm contribution} : W_R \ {\rm contribution} &=&
   \frac{g_L^2}{M_L^2}V^L_{cb}V^{L*}_{cs}(1+P):
   \frac{g_R^2}{M_R^2}V^R_{cb}V^{R*}_{cs}(1+P')   \nonumber \\
 &=& 1: \beta_g \frac{cse^{i(\sigma-\xi)}}{A\lambda^2} \frac{1+P'}{1+P} ,
\end{eqnarray}
where $P$ ($P'$) is $W_L$ ($W_R$) penguin contribution. We have
$P\sim P' \propto \alpha_S \ln (m_t^2/m_c^2)$
in the first approximation, so that we take $(1+P')/(1+P)=1$.
By putting
$\lambda=0.22$, $A=0.8$, $c=s=1/\sqrt{2}$ and $M_R>0.52$ TeV we find
$W_R$ contribution is 30 \% or less of the $W_L$ contribution,
which corresponds to $18^\circ$ or less
deviation of $\phi_1^{exp}$ from $\phi_1$. Note that the constraints
given in the previous section do not affect this conclusion except for
$W_R$ mass bound because $W_R$ does not contribute significantly
to $B^0$-$\overline{B^0}$ mixing and the solution (\ref{eqn:soli})
to suppress $\epsilon$ is independent of $\sigma-\xi$.

In the case of $V^R$ of type II there are no tree nor penguin
contributions by $W_R$ in this decay mode.
But there can be significant contribution to
$B^0$-$\overline{B^0}$ mixing which gives rise to
deviation of $\phi_1^{exp}$ from
the standard model value,
$\delta\phi_1=\delta\phi_{SM}/2=66^\circ \sim 115^\circ$,
as discussed in sec.2.3.2.
\subsection{$\phi_2$}
Measurement of the asymmetry in $B^0,\overline {B^0} \rightarrow \pi\pi$ gives
the angle $\phi_2^{exp}$. The involving quark mixing matrix elements are
$V_{ub}$ and $V_{ud}$. No tree nor penguin
$W_R$ contributions exist in
this decay mode in both cases of type I and type II. There is no significant
contribution to $B^0$-$\overline{B^0}$ mixing in the case of type I, so that
$\phi_2^{exp}$ coincides with  $\phi_2$. But large contribution is possible in
type II case, which leads to large deviation between
$\phi_2^{exp}$ and $\phi_2$ just as estimated
in the preceding subsection for  $\phi_1$.
\subsection{$\phi_3$}
The $CP$ angle $\phi_3^{exp}$ is fixed through the decays of $B$ mesons
into neutral $D$ ($D^0$, $\overline{D^0}$, $CP$ eigenstates of $D$) and $K$.
The involving quark mixing matrix elements are
$V_{cb}V_{us}^*$, $V_{ub}V_{cs}^*$ and their complex conjugates.
Penguin contribution is absent in this decay mode.
There is no $W_R$ contribution in
the case of type I, while there is a tree level contribution in the case of
type II.
\begin{eqnarray}
   W_L \ {\rm contribution} : W_R \ {\rm contribution} &=&
   \frac{g_L^2}{M_L^2}V^L_{cb}V^{L*}_{us}:
   \frac{g_R^2}{M_R^2}V^R_{cb}V^{R*}_{us}   \nonumber \\
 &=& 1: \beta_g \frac{e^{i(\sigma-\omega)}}{A\lambda^3}  .
\end{eqnarray}
To estimate the deviation $\delta\phi_3\equiv \phi_3^{exp} - \phi_3$
we define $\beta_g/(A\lambda^3)\equiv r_3$ and
replace $V^L_{cb}V^{L*}_{us}$ by $V^L_{cb}V^{L*}_{us}
[1+r_3 e^{i(\sigma-\omega)}]
=V^L_{cb}V^{L*}_{us}[1 \pm r_3 e^{i(\sigma-\varphi)}]$, where
eq.(\ref{eqn:sol}) has been
used. The discussion at sec.2.3.2 gives $\sigma-\varphi-\phi_{SM}
= 154^\circ \sim 206^\circ$ and $\phi_{SM}=2\phi_1=14^\circ \sim 30^\circ$
for  $m_t=175$ GeV, $r_B=1.3$ and $M_R=1.2$ TeV.
Then $\sigma-\varphi=168^\circ \sim 236^\circ$, and
we find $\delta\phi_3=-32^\circ \sim +19^\circ$ from Fig.5.
We have analyzed  $\delta\phi_3$ for other set of parameters. The results are
given in Table 3.

  \begin{center}
    \epsfig{figure=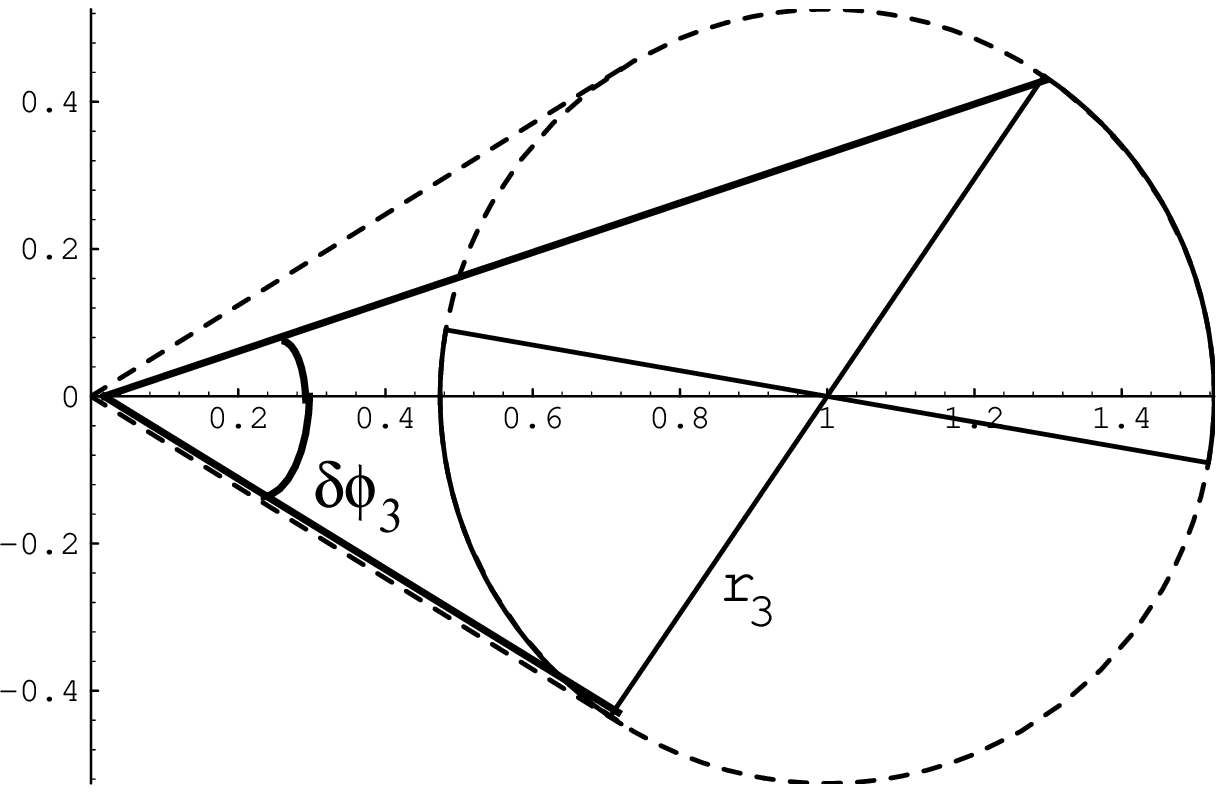,height=5cm}
\vspace*{4mm}

Figure 5:
\begin{minipage}[t]{10cm}
$\delta\phi_3$ for $r_B=1.3$ and $M_R=1.2$ TeV in the case of type II.
The solid parts of the circle express $r_3 e^{i(\sigma-\varphi)}$
with $\sigma-\varphi = 168^\circ \sim 236^\circ$ (mod $180^\circ$).
\end{minipage}
  \end{center}
  \begin{center}
  \begin{tabular}{c||ccc}
$r_B$ & $m_t$ = 150 (GeV) & 175 & 200 \\ \hline
1.3 &
\begin{tabular}{c}
$-32^\circ \sim +17^\circ$ ($M_R=1.2$) \\
$-11^\circ \sim +11^\circ$ ($M_R=2.0$)
\end{tabular}
 &
\begin{tabular}{c}
$-32^\circ \sim +19^\circ$ ($M_R=1.2$) \\
$-11^\circ \sim +11^\circ$ ($M_R=2.0$)
\end{tabular}
 &
\begin{tabular}{l}
$-37^\circ \sim -12^\circ$,  \\
$\ +3^\circ \sim +14^\circ$ ($M_R=1.1$) \\
$-11^\circ \sim +11^\circ$ ($M_R=2.0$)
\end{tabular}
 \\ \hline
1.0 &
\begin{tabular}{c}
$-23^\circ \sim +22^\circ$ ($M_R=1.4$) \\
$-11^\circ \sim +11^\circ$ ($M_R=2.0$)
\end{tabular}
  &
\begin{tabular}{l}
$-27^\circ \sim -10^\circ$,\\
$\ +4^\circ \sim +19^\circ$ ($M_R=1.3$) \\
$-11^\circ \sim +11^\circ$ ($M_R=2.0$)
\end{tabular}
 &
\begin{tabular}{l}
$-27^\circ \sim \ -5^\circ$,\\
$\ +2^\circ \sim +20^\circ$ ($M_R=1.3$) \\
$-11^\circ \sim +11^\circ$ ($M_R=2.0$)
  \end{tabular}
\end{tabular}

\vspace*{4mm}
 Table 3 : The values of $\delta\phi_3$.
The masses of $W_R$ boson ($M_R$) is given in unit of TeV.
    \end{center}

%
\section{Summary and discussion}
We have investigated the effects of $W_R$ on $CP$ violation in $B$ decays.
The right-handed charged current gauge boson
$W^R$ can affect significantly on the determination of the $CP$ violation
angles to be measured at $B$ factories if the right-handed current
quark mixing matrix $V^R$ is chosen to satisfy the bounds by neutral $K$
meson system with relatively light $W_R$ of $M_R=0.5 \sim 1.5$ TeV. The
$W_R$ contribution can be sizable in $B^0$-$\overline{B^0}$ mixing and
tree level $b$ quark decay.
In the case of $V^R$ of type I the $CP$ angle $\phi_1$ ($\beta$) can
deviate from the standard model value by as large as $18^\circ$ for
$M_R = 0.52$ TeV. In the case of $V^R$ of type II
the $CP$ angle $\phi_1$ ($\beta$) and $\phi_2$ ($\alpha$) can
deviate by $66^\circ \sim 115^\circ$  and $\phi_3$ ($\gamma$)
by $-32^\circ \sim +19^\circ$ for $M_R = 1.2$ TeV. These results have
been obtained under specific sets of parameters for the simplicity of
calculation;
$A=0.8$, $\sqrt{(1-\rho)^2+\eta^2}=1.0, 1.3$, $B_K=0.6 \sim 1.0$,
$B_B=0.7 \sim 1.1$, $g_L=g_R$ and neglecting $W_L$-$W_R$ mixing.
The values of deviation will be modified if we enlarge or restrict
the region of these parameters.

One notable point of the $W_R$ effects is the fact that
the sum of the measured
three $CP$ angles does not become $180^\circ$. It has often been
pointed out that a key of new physics search in $B$ meson system is
a check of sum of three $CP$ angles\cite{newb}. However, the sum of
the angles measured on $\Upsilon (4S)$, where the $CP$ angle $\phi_3$
($\gamma$)
is fixed through $B \rightarrow DK$ decay,  becomes $180^\circ$  even
with sizable new physics contributions if new physics affects
on $B^0$-$\overline{B^0}$ mixing alone as can be seen in the definitions
of the $CP$ angles to measure. This result can be extended to 4 or more
generation models, vector-quark models and so on where Kobayashi-Maskawa
matrix is not necessarily unitary in the first $3\times3$ part\cite{kt}.
$SU(2)_L \times SU(2)_R \times U(1)$ models will be one of
promising candidates of new physics if sum of the three $CP$ angles
measured in coming experiments does not become $180^\circ$.

\vspace*{3cm}
\begin{center}
  \Large{Acknowledgement}
\end{center}

\noindent
The authors would like to thank Dr. M.~Tanaka for useful comments
on QCD correction factors in $SU(2)_L \times SU(2)_R \times U(1)$ models.
T.K.'s work is supported in part
by Grant-in Aids for Scientific Research from the Ministry
of Education, Science and Culture (No. 07804016 and 07304029).



\begin{thebibliography}{99}
\bibitem{KEKB} {\it Letter of Intent for a Study of CP
Violation in B Meson Decays}, KEK Report 94-2 (1994).
\bibitem{SLACB} {\it Letter of Intent for the Study of CP
Violation and Heavy Flavor Physics at PEP II}, SLAC-443 (1994).
\bibitem{unit} A.B.~Carter and A.I.~Sanda,
Phys. Rev. Lett. {\bf 45} 952 (1980),
Phys. Rev. {\bf D23} 1567 (1981); \\
I.I.~Bigi and A.I.~Sanda, Nucl. Phys. {\bf B193} 85 (1981),
Nucl. Phys. {\bf B281} 41 (1987).
\bibitem{newb} For reviews see \\
Y.~Nir, {\it Proc. of the Workshop on B physics at Hadron Accelerators},
eds. P.~McBride and C.S.~Mishra, SSCL-SR-1225,
Fermilab-CONF-93/267, 185 (1993);\\
Y.~Nir and H.R.~Quinn, Ann. Rev. Nucl. Part. Sci. {\bf 42} 211 (1992).
\bibitem{KM} M.~Kobayashi and T.~Maskawa, Prog.Theor.Phys. {\bf 49},
652 (1973).
\bibitem{lrmd} R.N.~Mohapatra and J.C.~Pati, Phys. Rev. {\bf D11} 566 (1975),
{\bf D11} 2558 (1975);\\
G.~Senjanovic and R.N.~Mohapatra, Phys. Rev. {\bf D12} 1502 (1975);
G.~Senjanovic, Nucl. Phys. {\bf B153} 334 (1979).
\bibitem{mrcon} G.~Beall, M.~Bander and A.~Soni, Phys. Rev. Lett.
{\bf 48} 848 (1982).
\bibitem{OE} F.I.~Olness and M.E.~Ebel, Phys. Rev. {\bf D30} 1034 (1984).
\bibitem{LS} P.~Langacker and S.U.~Sankar, Phys. Rev. {\bf D40} 1569 (1989).
\bibitem{LW} D.~London and D.~Wyler, Phys. Lett. {\bf B232}  503 (1989).
\bibitem{PDG} L.~Montanet et.al., Phys. Rev. {\bf D50} 1173 (1994)
and 1995 off-year partical updata for 1996 edition available on
the PDG WWW pages (URL:http://pdg.lbl.gov/).
\bibitem{gamma}  M.~Gronau and D.~London, Phys. Lett. {\bf B253} 483 (1991);\\
M.~Gronau and D.~Wyler,  Phys. Lett. {\bf B265} 172 (1991);\\
I.~Dunietz, Phys. Lett. {\bf B270} 75 (1991).
\bibitem{ecg} G.~Ecker and W.~Grimus, Nucl. Phys. {\bf B258} 328 (1985);\\
H.~Nishiura, E.~Takasugi and M.~Tanaka, Prog. Theor. Phys.
{\bf 84} 116 (1990),
Prog.Theor.Phys.{\bf 84} 502 (1990),
Prog.Theor.Phys.{\bf 85} 343 (1991).
\bibitem{datta} A.~Datta, E.A.~Paschos, J.-M.Schwartz and M.N.Sinha Roy,
hep-ph/9509420 (1995), DO-TH 95/12.
\bibitem{wolf} L.~Wolfenstein, Phys. Rev. Lett. {\bf 51} 1945 (1983).
\bibitem{blow} T.E.~Browder and K.~Honscheid, UH 511-816-95,
OHSTPY-HEP-E-95-010 (1995).
\bibitem{kt} T.~Kurimoto and A.~Tomita, in preparation.
\end{thebibliography}
\end{document}